\documentclass[conference]{IEEEtran}
\IEEEoverridecommandlockouts
\usepackage{cite}
\usepackage{amsmath,amssymb,amsfonts}
\usepackage{algorithmic}
\usepackage{graphicx}
\usepackage{textcomp}
\usepackage{xcolor}
\def\BibTeX{{\rm B\kern-.05em{\sc i\kern-.025em b}\kern-.08em
    T\kern-.1667em\lower.7ex\hbox{E}\kern-.125emX}}
\begin{document}

\title{Self-Relation Attention and Temporal Awareness for
           Emotion Recognition via Vocal Burst \\ 
\thanks{Correspondence to: Guee-Sang Lee (gslee@jnu.ac.kr)}
}

\author{\IEEEauthorblockN{1\textsuperscript{st} Dang-Linh Trinh}
\IEEEauthorblockA{\textit{Department of AI Convergence} \\
\textit{Chonnam National University)}\\
Gwangju, Korea \\
linhtd812@gmail.com}
\and
\IEEEauthorblockN{2\textsuperscript{nd} Minh-Cong Vo}
\IEEEauthorblockA{\textit{Department of AI Convergence} \\
\textit{Chonnam National University}\\
Gwangju, Korea \\
congvm.it@gmail.com}
\and
\IEEEauthorblockN{3\textsuperscript{rd} Guee-Sang Lee $^*$}
\IEEEauthorblockA{\textit{Department of AI Convergence} \\
\textit{Chonnam National University}\\
Gwangju, Korea \\
gslee@jnu.ac.kr} \\
}

\maketitle

\begin{abstract}
The technical report presents our emotion recognition pipeline for high-dimensional emotion task (A-VB High) in The ACII Affective Vocal Bursts (A-VB) 2022 Workshop \& Competition. Our proposed method contains three stages. Firstly, we extract the latent features from the raw audio signal and its Mel-spectrogram by self-supervised learning methods. Then, the features from the raw signal are fed to the self-relation attention and temporal awareness (SA-TA) module for learning the valuable information between these latent features. Finally, we concatenate all the features and utilize a fully-connected layer to predict each emotion's score. By empirical experiments, our proposed method achieves a mean concordance correlation coefficient (CCC) of 0.7295 on the test set, compared to 0.5686 on the baseline model. The code of our method is available at https://github.com/linhtd812/A-VB2022.
\end{abstract}


\begin{figure*}[ht]
\centering
\label{fig1}
\includegraphics[width=0.8\textwidth]{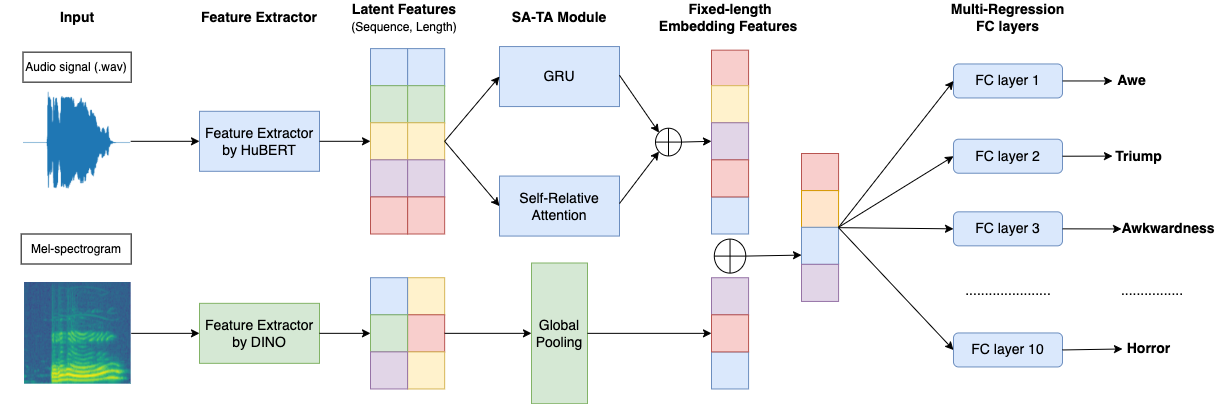}
\caption{Overall Architecture of our proposed method}
\end{figure*}

\section{Introduction}
Human speech is one of the most valuable resources to help identify people's emotions or feelings \cite{b7}. There are several kinds of human speech, including verbal and non-verbal. Recently, tremendous research has been conducted in speech emotion recognition with verbal speech, which is applied in human-computer interfaces. Non-verbal speech, known as vocal burst (VB), is the voice signal without meaning by a human being but could be translated to words such as laughter, groans, and grunts. However, the research on VB field is sparse because of the lack of data related to non-verbal human speech. Therefore, to discover the new trend of speech emotion recognition (SER), A-VB 2022 competition \cite{b2} provides us with the HUME-VB corpus \cite{b1} to find the meaning of VB related to people's emotions. For example, laughter could have some related emotion like amusement or triumph.    

In supervised learning, data augmentation could enlarge the scale of data for over-fitting prevention and model generalization improvements \cite{b8}. Besides, self-supervised learning \cite{b14} is a trending method, which could learn the generic representation from large-scale data without manual annotations. From this point, recent research proved that the SSL model could achieve competitive results compared to the supervised learning method \cite{b16}. Also, the pre-trained SSL model is utilized for feature extracting in many downstream applications \cite{b9}. Attention mechanisms significantly impact deep learning models in many fields, which enrich the information the model could learn from inputs \cite{b15}. Attention mechanisms can select, modulate, and focus on the information most important to the target of our problem, like human attention \cite{b10}. Therefore, this paper will investigate the effectiveness of data augmentation, SSL models, and attention modules on emotion recognition via vocal burst.

This technical report focuses on the first task of the A-VB 2022 competition: the High-Dimensional Emotion Task (A-VB High). This task's target is to predict the scores of 10 emotions. In our proposed method, several techniques and our contribution are listed below.

\begin{itemize}
\item We investigate the efficiency of self-supervised learning (SSL) for extracting the latent features from both raw audio signal and its Mel-spectrogram by applying HuBERT \cite{b6} and DINO \cite{b7} models.
\item The Self-Relation Attention and Temporal Awareness (SA-TA) module helps captures the meaningful information from not only essential parts in audio signal but also the temporal information of latent features extracted from HuBERT \cite{b6} model. 
\item The result improves slightly by utilizing a Mel-spectrogram containing the information related to the frequency and loudness of VB.
\end{itemize}

The paper contains the list of content as follows. Section 2 describes the data analysis and pre-processing of this data. After that, the architecture of the proposed method is described in detail in section 3. Experimental results are shown in section 4, and section 5 concludes the overall paper.

\section{Data Analysis and Pre-processing}

The Hume Vocal Burst Database (H-VB) \cite{b1} is utilized for the ACII A-VB 2022 challenge, which consists of 59,201 non-vocal audio from 1,702 speakings from 4 different cultures, including the U.S, South Africa, China, and Venezuela. Also, the dataset is split into the train, validation, and test subsets. The labels for the A-VB High task are the scores for each emotion, and we evaluate the results based on the mean CCC metric over ten emotion scores. There are ten basic emotions for the A-VB High task: Awe, Excitement, Amusement, Awkwardness, Fear, Horror, Distress, Triumph, Sadness, and Surprise.

There are two audio forms, including .wav and .webm files (a compressed format). Our team utilizes the .wav format with a sample rate of 16kHZ. For pre-processing, we trim the silence in the audio file and set the duration of the audio input is 3.5 seconds. Then, we apply some augmentation techniques to a raw audio signal, including random pitch shift and random time warping, to enlarge the scale of the data. If the duration after trimming is smaller than 3.5 secs, we add zero-padding at the beginning of the audio file. Otherwise, we randomly cut the sample into an audio file with a duration of 3.5 secs. Also, after applying the pre-processing, we transform the processed audio signal to Mel-spectrogram as the other input of our proposed method.

\begin{table*}[htbp]
\caption{The mean CCC on validation dataset from different models}
\begin{center}
\scalebox{1.1}{
\begin{tabular}{|l|c|}
\hline
\textbf{Model} & \textbf{\textit{Mean CCC}}\\
\hline
Baseline \cite{b2} & 0.5638 \\
Wav2Vec2-large & 0.6902\\
HuBERT-large & 0.7012\\
DINO & 0.5920 \\
HuBERT-large + SA-TA & 0.7265\\
HuBERT-large + DINO + SA-TA & 0.7303\\
\hline
\end{tabular}}
\label{tab1}
\end{center}
\end{table*}

\begin{table*}
\caption{Evaluation on each emotion on validation and test dataset based on CCC metric of our proposed method}
\centering
\label{tab2}
\scalebox{1}{
\begin{tabular}{|l|cccccccccc|c|}
\hline
\textbf{Dataset} & \textbf{Awe} & \textbf{Excite.} & \textbf{Amuse.} & \textbf{Awkward.} & \textbf{Fear} & \textbf{Horror} & \textbf{Distress} & \textbf{Triumph} & \textbf{Sadness} & \textbf{Surprise} & \textbf{Mean CCC} \\
\hline
Validation & 0.8084 & 0.6895 & 0.7886 & 0.6080 & 0.7614 & 0.7370 & 0.6959 & 0.6813 & 0.7069 & 0.8125 & 0.7303 \\
Test & 0.8140 & 0.6817 & 0.7956 & 0.6100 & 0.7623 & 0.7362 & 0.6935 & 0.6778 & 0.7128 & 0.8113 & 0.7295\\
\hline
\end{tabular}}
\end{table*}

\section{Methodology}
In this section, we describe the architecture of our proposed method. Firstly, we introduce our approach generally. After that, each component of this pipeline will be described in detail.

\subsection{Architecture Overview}
Our proposed method is shown in Fig. 1. The input for this architecture is the pre-processed audio waveform and the Mel-spectrogram (mentioned in the previous section). Then, the self-supervised learning method includes Hidden-Unit Bert \cite{b6}, and DINO \cite{b7} for extracting the latent features from original inputs. While HuBERT is applied for the audio signal, Mel-spectrogram is treated as images, and DINO is utilized to extract features from these input types. After that, latent features extracted from HuBERT are fed into an SA-TA module to accentuate the vital time point in each audio signal. Finally, we concatenated all the features and fed them in FC layers to predict the score of each emotion individually.

\subsection{Features Extraction by Self-Supervised Learning Method}
Self-supervised learning (SSL) is the method that learns from unlabeled sample data. Recently, SSL has been utilized as a pre-task to learn nontrivial data representations. Inspired by [1], we explore two pre-trained SSL models, HuBERT \cite{b6} for audio signal and DINO \cite{b7} for its Mel-spectrogram. For audio signal, we hypothesize that HuBERT can capture the general information, not only acoustic information, but also the phonemes of VB. Besides, by utilizing the Mel-spectrogram, we capture helpful information on the frequency and loudness of the sound. Furthermore, by using pre-trained models on large-scale dataset like Hubert and DINO, we can fine-tune, which lead to better latent features for the following stages in our proposed method. All the pre-trained models are based on the Transformer achitecture. While HuBERT model is pre-trained on the Libri-light dataset \cite{b17} for speech recognition without supervision, DINO pre-trained weights on the Google Landmark v2 dataset \cite{b18} are utilized for extracting features from the Mel-Spectrogram of audio signals. 

\subsection{Self-Relation Attention and Temporal Awareness Module}
The Self-Relation Attention and Temporal Awareness (ST-TA) module consists of two parts, including Self-Relation Attention, which teaches self-attention for each feature and the relationship between all the time-point features. Self-Relation Attention is inspired by \cite{b3}. We hypothesize that this could automatically capture the vital part of each latent feature because the vocal burst is concise, and the meaningful information only appears for a short time, not all the duration of an audio sample. Furthermore, we can consider latent features extracted from HuBERT \cite{b6} as time-series data so that the Gated Recurrent Unit (GRU) \cite{b4} is considered a Temporal Awareness Part for capturing temporal information of these features.

\subsection{Multi-Labels Regression Module}
The latent features from DINO \cite{b7} go through the Global Pooling module to be a one-dimensional vector. After that, we concatenated all the features from the previous module and fed them into fully-connected (FC) layers. Because of the multi-regression problem, we use ten separate FC layers to predict the score of each emotion. The score of each emotion is a range of (0,1). 

\subsection{Loss Function}
Because all the test results are evaluated by Concordance Correlation Coefficient (CCC) metric, our loss function is designed based on the CCC metric below. CCC is the concordance between prediction (1) and the ground truth (2), which identifies the agreement between two variables.

\begin{displaymath}
L_{CCC} = 1 - \frac{2 \sigma_{12}}{(\mu_1-\mu_2)^2 + \sigma_1^2 + \sigma_2^2}
\end{displaymath}

where $\smash{\sigma}$, $\smash{\mu}$ and $\smash{\sigma_{12}}$ are denoted by mean, standard deviation and their covariance between predicted score and ground truth, respectively.

\section{Experimental Results}
\subsection{Experimental Setup}
As input features, we use both raw audio signal and its Mel-spectrogram. Some function in the Torchaudio package augments each raw audio signal. Through the experiments, the Adam optimizer is applied with a learning rate of 1e-5, and early stopping is utilized with an patience of 10 epochs to prevent over-fitting. Also, the learning rate is halved if the loss on the validation dataset does not decrease. The maximum epochs of the training process are set to 50, and the batch size is 4. All the model is trained with Nvidia RTX 2080Ti GPU and Pytorch 1.7.1.
\subsection{Experimental Results}
Firstly, we investigate the efficiency of recent SSL models for audio signal including Wav2vec2-large \cite{b11} and HuBERT-large \cite{b6}. These two models were trained on public large dataset such as Libri-Light and Librispeech.

From Table \ref{tab1}, the HuBERT model is better for non-verbal emotion recognition tasks than Wav2Vec2. Also, the large version of SSL models is chosen because the effectiveness is illustrated in previous research \cite{b13}. 

Non-verbal speech is always expressed in a short duration. Therefore, the valuable information is only in a short time or several time-point during the audio signal. Therefore, we use the SA-TA module to help the model focus on valuable time-point from latent features extracted from HuBERT. Using this module, the average CCC on the validation dataset increases by 0.2 compared to using only the HuBERT model for extracting features. The result of the SA-TA module is shown in Table \ref{tab1}.

Finally, by using DINO for extracting features from Mel-spectrogram and global pooling module, we get slightly better results than the baseline model, which achieves 0.5920 of the mean CCC metric. However, using the feature from the DINO model is not good compared to those from HUBERT because the dataset for DINO is in another domain, not trained in the speech signal domain like the HUBERT pre-trained model. By combining both features from DINO and HUBERT model, we find that the improvement of mean CCC by up to around 0.005 compared to using only raw audio signal. The result shows that the information about the frequency and the loudness of the sound is valuable for emotion prediction related to non-verbal human speech.

We evaluate the CCC metric on each emotion by applying the proposed method, shown in Table \ref{tab2}. From the empirical experiment, our method works better on Awe and Surprise emotions than others. Furthermore, the result on the test dataset is 0.7295, almost the same as the validation dataset, which means the model has a good generalization ability.

\section{Conclusion}
The technical report shows the emotion recognition method for vocal burst submitted to the High-Dimensional Emotion Task of A-VB 2022 challenge. The proposed architecture use SSL models to extract latent feature from raw signal and its Mel-spectrogram. ST-TA module is applied to focus on the critical time-point from latent features and use multi-regression FC layers to individually predict scores of ten emotions. Experiment results show that our proposed method achieves 0.7295 mean CCC compared to 0.5686 of the baseline model from the competition organizer. 

\section*{Acknowledgment}

This research was supported by the Bio \& Medical Technology Development Program of the National Research Foundation (NRF) \& funded by the Korean government (MSIT) (NRF-2019M3E5D1A02067961) and by Basic Science Research Program through the National Research Foundation of Korea (NRF) funded by the Ministry of Education (NRF-2018R1D1A3B05049058 \& NRF-2020R1A4A1019191).

\vspace{12pt}

\end{document}